\documentclass[aps,twocolumn,showpacs,amsmath,amsfonts,amssymb]{revtex4}
\usepackage{graphicx}
\usepackage{epsf}

\begin{document}
\title{Sub-Planck scale structures in a vibrating molecule in the presence of decoherence}
\author{Suranjana Ghosh$^{\mathrm{1}}$\footnote{e-mail:
suranjana.ghosh@unicam.it}, Utpal
Roy$^{\mathrm{1}}$\footnote{e-mail: utpal.roy@unicam.it}, Claudiu
Genes$^{\mathrm{2}}$\footnote{e-mail: claudiu.genes@unicam.it} and
David Vitali$^{\mathrm{1}}$\footnote{e-mail:
david.vitali@unicam.it}}

\affiliation{$^{\mathrm{1}}$Dipartimento di Fisica, Universit\`{a}
di Camerino, I-62032 Camerino, Italy\\$^{\mathrm{2}}$Institute for
Theoretical Physics, University of Innsbruck, A-6020 Innsbruck,
Austria}

\begin{abstract}
We study the effect of decoherence on the sub-Planck scale
structures of the vibrational wave packet of a molecule. The time
evolution of these wave packets is investigated under the
influence of a photonic or phononic environment. We determine the
master equation describing the reduced dynamics of the wave-packet
and analyze the sensitivity of the sub-Planck structures against
decoherence in the case of a hydrogen iodide (HI) molecule.

\end{abstract}
\pacs{42.50.Md, 03.65.Yz, 33.80.-b} \maketitle
%
%
%%%%%%%%%%%%%%%%%%%%%%%%%% MY CONTENTS %%%%%%%%%%%%%%%%%%%%%%%%%%
\section{Introduction}
Recent progress of controlled femtosecond pulses has advanced
greatly the technology during the last few years \cite{zewail1}. A
new field of molecular optics has emerged where lasers are used to
manipulate the internal and external degrees of freedom of
molecules, to deflect beams of molecules, to control molecular
dynamics, and to align molecules \cite{shapiro,garraway}. Many
investigations have focused on the vibrational motion of diatomic
molecules. The single bond between the atoms acts as a spring and
supports harmonic oscillations for small amplitudes, but the bond
can break (dissociate) when stretched too much. These phenomena
usually occur at time scales between few picoseconds and few
hundred femtoseconds. With ultrashort pulses one can now prepare a
molecular wave packet and probe its evolution and observe
molecular reactions in this time domain. Successful experiments
have been performed on several molecules \cite{expt}. The most
convenient model for studying the vibrational motion of diatomic
molecule is the Morse potential, which is an exactly solvable
system \cite{morse}. Coherent superposition of several vibrational
levels of the molecule creates the wave packet which, due to
quantum interference, shows revival and fractional revivals
\cite{averbukh,bluhm,robinett} in their time evolution.

Fractional revivals are associated with superpositions of
separated wave packets (for example, the so-called Schr\"odinger
cat states), which manifest clear quantum interference effects and
nonclassical features, which can be well visualized in the phase
space of the vibrational motion. Although the experimental
observation of small quantum interference structures is very
challenging, it has already been visualized in Pico-meter scale
\cite{ohmri}. A number of different phase space distribution
functions have been introduced \cite{cahill} and investigated over
the years, and among these the Wigner distribution \cite{schleich}
is particularly useful, because its negativity yields an
indication of nonclassical behavior \cite{foldiopt,vogel1,vogel2}.
Zurek \cite{zurek2} first showed that this negativity reveals the
existence of the smallest structures in phase space i.e., the
sub-Planck scale structures. One may expect that Heisenberg's
uncertainty principle implies that structures on scales smaller
than the Planck constant have no observable consequence, while
instead  Zurek \cite{zurek2} showed that these highly nonclassical
structures are expected to be particularly sensitive to
decoherence. Through a short walk in controversy, recently
sub-Planck scale structures draw considerable attention and have
been found by others in different situations
\cite{ph,wisn,pathak1,pathak2,toscano,dalvit,praxmeyer,ghosh,manan,fibre,jay,scott}.

Decoherence due to the coupling to an external environment is the
main responsible for the disappearance of nonclassical
manifestations of quantum states and it is considered one of the
mechanisms through which the classical world at the macroscopic
level emerges from the quantum substrate~\cite{zurek1,zoller,zeh}.
Decoherence on the molecular vibrational degree of freedom is due
to the coupling between vibrational and rotational modes
\cite{brif} and also to the coupling with the photonic and
phononic degrees of freedom~\cite{foldi}. The latter are
associated with a super-Ohmic environment describable in terms of
a continuous set of bosonic modes and in this paper we shall focus
on the effect of this source of decoherence. To be more specific,
we will study the sub-Planck scale structures in the Morse system
and the effect of decoherence on these structures in molecular
wave packets. We shall determine the master equation describing
the reduced dynamics of the wave packet and analyze the robustness
of the sub-Planck structures against decoherence.

The paper is organized as follows. In Sec.~II we give a brief
overview of the Morse potential and its coherent states, while in
Sec.~III we derive the master equation in the case of the coupling
with a bosonic environment at thermal equilibrium. In Sec.~IV we
study the effect of decoherence on the Wigner function at the
sub-Planck level and the sensitivity to decoherence of these
structures is analyzed. Finally, we conclude in Sec.~V.

\section{Review of the Morse model of a vibrating molecule}

Vibrational dynamics of diatomic molecules are well described by
Morse potential \cite{morse,vetchin1,vetchin2,dahl,wolf}. It can
be described as
\begin{equation}
V(x)=D(e^{-2\beta x}-2e^{-\beta x})
\end{equation}
 where
$x=r/r_{0}-1,\;r_{0}$ is the equilibrium value of the
inter-nuclear distance $r$ and $\beta$ is a range parameter. $D$
is the dissociation energy, which has been extensively studied in
a wider context of this model \cite{robert,benedict}. Defining
\begin{equation}
\lambda=\sqrt{\frac{2\mu D
r^{2}_{0}}{\beta^2\hbar^2}}\;\;\mathrm{and}\;\;
s=\sqrt{-\frac{8\mu r^{2}_{0}}{\beta^2\hbar^2}E},
\end{equation}
where $\mu$ is the reduced mass of the vibrational motion, the eigenfunctions of the Morse potential can be written as
\begin{equation}\label{eigenstate}
\psi_{n}^{\lambda} (\xi)= N e^{-\xi/2} \xi^{s/2} L_{n}^{s} (\xi),
\end{equation}
where $\xi=2\lambda e^{-\beta x}$, $0<\xi<\infty$, and
$n=0,1,...,[\lambda-1/2]$, with $[\rho]$ denoting the integer part
of $\rho$, so that the total number of bound states is
$[\lambda-1/2]$. The parameters $\lambda$ and $s$ satisfy the
constraint condition $s+2n=2\lambda-1$.

Note that $\lambda$ is potential dependent and $s$ is related to
energy $E$ and, by definition,  $\lambda>0,\;s>0$. In
Eq.~(\ref{eigenstate}),
 $L_{n}^{s}(y)$ is the associated Laguerre polynomial and $N$ is the
normalization constant:
\begin{equation}
N=\left[\frac{\beta(2\lambda-2n-1)\Gamma{(n+1)}}{\Gamma{(2\lambda-n)}r_{0}}\right]^{1/2}.
\end{equation}
Revival and fractional revivals appear during the time evolution
of a suitably prepared wave packet and are well studied in the
literature \cite{averbukh,bluhm,robinett}. Here we study the
effect of decoherence on the motion of a molecular wave packet
through its sub-Planck scale structures. There structures are
found at one eighth of the fractional revival time in the Wigner
phase space distribution \cite{ghosh}. The initial wave packet is
taken here as SU$(1,1)$ coherent state (CS) of this potential
\cite{kais}, which is obtained upon applying the displacement
operator on the ground state. The CS is given by
\begin{eqnarray}
|\eta,s\rangle&=&e^{\alpha K_{+}-\alpha^{*}K_{-}}|0,s\rangle\nonumber\\
&=&(1-|\eta|^{2})^{\frac{1+s}{2}}\sum^{\infty}_{k=0}\left[\frac{\Gamma(k+s+1)}{k!\Gamma(1+s)}\right]^{1/2} \eta^{k}|k,s\rangle. \label{cs}
\end{eqnarray}
where $0\leq k \leq[\lambda-1/2]$ correspond to the bound states
of the Morse potential and $k>[\lambda-1/2]$ are the appropriate
scattering states. The parameter $\eta$ is associated with the
``amplitude'' of the CS and possesses the same phase of the
displacement amplitude $\alpha$, while its modulus is given by the
relation $|\eta| =\tanh |\alpha|$~\cite{kais}. In our numerical
analysis, we will always consider low energy coherent states well
below the dissociation limit so that only the bound states of the
Morse potential can be used as basis set.
\begin{figure*}[htbp]\centering
  \includegraphics[width=5.in]{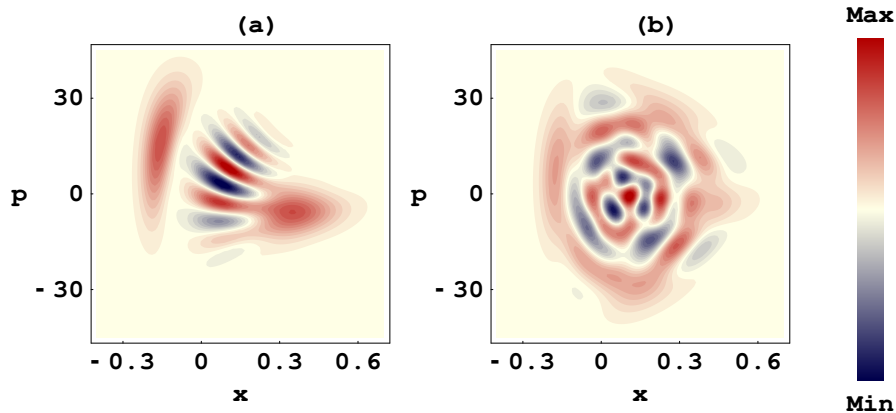}
\caption{(Color online) Time evolution of Morse wave packet in
phase space Wigner distribution: (a) cat state at time
$t=\frac{1}{4} T_{rev}$ and (b) sub-Planck scale structures,
appeared at the middle at time $t=\frac{1}{8} T_{rev}$, where
$\alpha=0.3$, $\beta=2.07932$, $r_{0}=3.04159$ a.u., and
$\bar{n}=4$. Here, $x$ and $p$ are the dimensionless position and
momentum variables, where $x=r/r_{0}-1$ and $p$ is the
corresponding scaled variable.} \label{catsub}
\end{figure*}

\section{Master equation for the Morse oscillator}

As described in Sec. I, we now investigate the effect of the
decoherence of an external phononic or photonic environment on the
sub-Planck scale structure. Therefore, the total model Hamiltonian
is \cite{zoller}
\begin{equation}
H=H_{sys}+H_{E}+H_{I},
\end{equation}
where $H_{sys}$ is the Morse Hamiltonian of the vibrational mode, $H_{E}$ is the environment Hamiltonian described by a set of independent
bosonic modes
\begin{equation}
H_{E}=\sum_{k}\hbar \omega_{k}(a^{\dagger}_{k}a_{k}+1/2),
\end{equation}
and $H_{I}$ is the interaction between the Morse particle and the environment, which we choose of the following form (see also
\cite{foldi,brumer})
\begin{equation}
H_{I}=\hbar \hat{O}^{\dagger}\sum_{k}\sigma_{k}a_{k}+ H.c.,
\end{equation}
where $\sigma_{k}$ are coupling constants. This choice corresponds
to assume the rotating wave approximation (RWA) in the interaction
with the environment so that we neglect counter-rotating terms,
while the operator $\hat{O}$ is a generic operator of the
vibrational mode, whose specific form depends on the considered
environment. Using standard techniques \cite{zoller}, one gets in
the usual Born-Markovian approximation, the following master
equation for the reduced density operator of the Morse oscillator
$\rho$,
\begin{eqnarray}
\frac{d}{dt}\rho &=&-\frac{i}{\hbar}[H_{sys},\rho] \label{meq}\\
&+&\left[\hat{O}_{2} \rho \hat{O}^{\dagger}+\hat{O} \rho \hat{O}_{2}^{\dagger} -
\hat{O}^{\dagger} \hat{O}_{2}\rho -\rho \hat{O}_{2}^{\dagger}\hat{O}\right] \nonumber \\
&+&\left[\hat{O}_{1}^{\dagger}\rho \hat{O}+\hat{O}^{\dagger} \rho \hat{O}_{1} - \hat{O}\hat{O}_{1}^{\dagger} \rho -\rho
\hat{O}_{1}\hat{O}^{\dagger}\right], \nonumber
\end{eqnarray}
where the operators $\hat{O}_j$ $(j=1,2)$ are new operators of the
vibrational mode corresponding to ``modifications'' of the
operator associated to the absorption from the environment
($\hat{O}_1$) or emission into the environment ($\hat{O}_2$) of
vibrational quanta. This fact is easily understood if we look at
their expression in the energy eigenbasis $|n,s\rangle$ used in
Eq.~(\ref{cs}). In fact, one has
\begin{equation}\label{ojs}
    \hat{O}_j=\sum_{m,n}O_{j}^{m,n}|m,s\rangle \langle n,s|,
\end{equation}
where
\begin{eqnarray}\label{elemat1}
 O_{1}^{m,n} &=& O^{m,n}\pi g(\omega_{mn})|\sigma(\omega_{mn})|^2 \bar{n}(\omega_{mn}),\\
 O_{2}^{m,n} &=& O^{m,n}\pi g(\omega_{mn})|\sigma(\omega_{mn})|^{2} \left[\bar{n}(\omega_{mn})+1\right]. \label{elemat2}
\end{eqnarray}
The quantities $O^{m,n}$ are the matrix elements of $\hat{O}$,
$g(\omega_{mn})$ is the density of states at the transition
frequency between two energy levels,
$\omega_{mn}=(E_m-E_n)/\hbar$, and
$\bar{n}(\omega_{mn})=\left[\exp\{\hbar\omega_{mn}/k_B
T\}-1\right]^{-1}$ is the mean thermal number of environmental
excitations, being the latter at equilibrium at temperature $T$.
The appearance of these two operators is a direct consequence of
the nonlinearity of the molecular vibrational motion. In fact, in
the linear case the transition frequencies $\omega_{mn}$ do not
depend on $n$ and $m$ and therefore $\hat{O}_1$ and $\hat{O}_2$
becomes proportional to $\hat{O}$. As a consequence, master
equation (\ref{meq}) become identical to the master equation of a
harmonic oscillator in a thermal environment in the RWA
\cite{zoller}.

\section{Sub-Planck scale structure and its sensitivity through decoherence}

We now solve the master equation (\ref{meq}) for the specific case
of the HI molecule and we adopt the Wigner function picture in
order to look at the effects of decoherence on sub-Planck scale
structures in phase space. The Wigner distribution is defined
($\hbar=1$) by
\begin{eqnarray}
W(x,p,t)=\frac{r_0}{2\pi}\int_{-\infty}^{\infty}\langle
x-\frac{x'}{2}|\rho(t)| x+\frac{x'}{2}\rangle e^{ix'p}dx',
\end{eqnarray}
and well describes the nonclassical interference effects associated with the time evolution of a wave packet in the nonlinear potential of the
Morse oscillator.

\begin{figure*}[htbp]\centering
  \includegraphics[width=2.2in,height=1.8in]{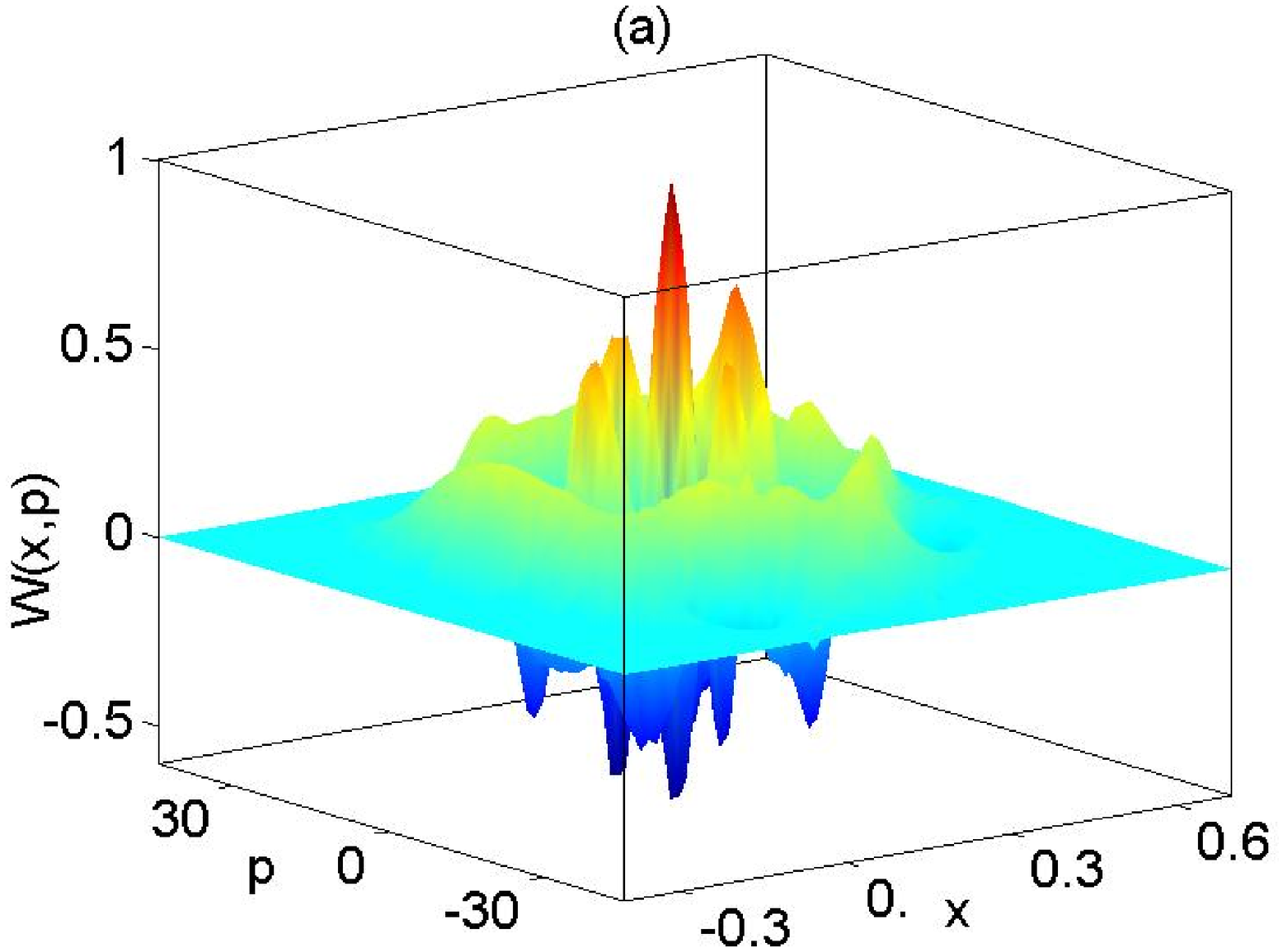}\!\!\!\!\!
  \includegraphics[width=2.2in,height=1.8in]{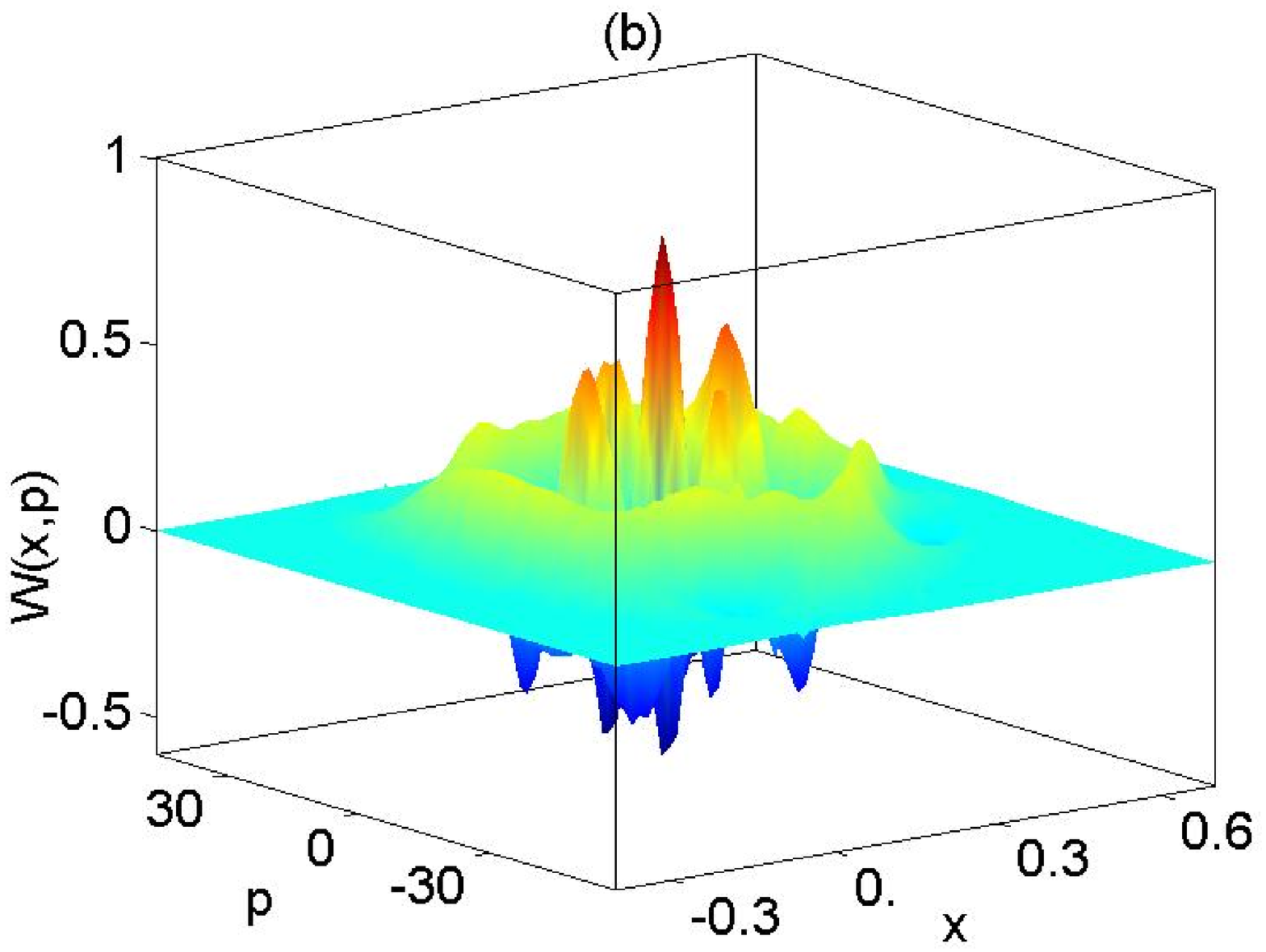}\!\!\!\!\!
   \includegraphics[width=2.6in,height=1.8in]{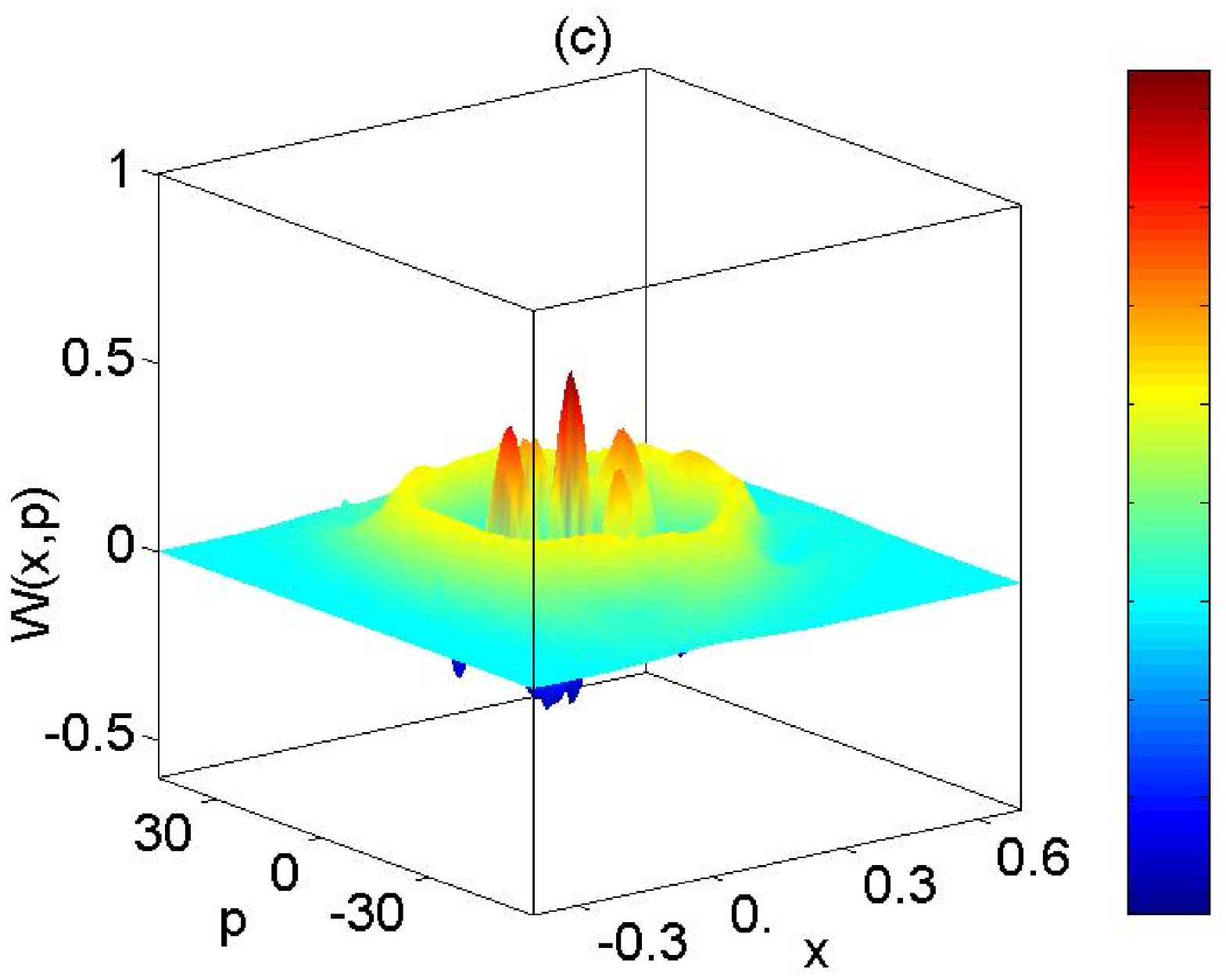}
\caption{(Color online) Wigner function of the coherent state at
one eighth of revival time, approximately equal to a compass
state, for three different values of the decoherence parameter
$\delta$ (a) $\delta=0$; (b) $\delta=0.54\times10^{3}$ a.u.; and
(c) $\delta=2.2\times10^{3}$ a.u. Here, $x$ and $p$ are the
dimensionless position and momentum variables, where $x=r/r_{0}-1$
and $p$ is the corresponding scaled variable. The environment
temperature is fixed at $T=10 \hbar\omega_{01}/k_B$.}
\label{3ddeco}
\end{figure*}

This fact is visible, for example, in Fig.~\ref{catsub}, which
shows the time evolution of an initial CS wave packet in phase
space at two different fractional revival times in the absence of
decoherence. We have considered a HI molecule, which has $30$
bound states, with $\beta=2.0793$, reduced mass $\mu=1819.99$
a.u., $r_{0}=3.0416$ a.u., and $D=0.1125$ a.u. \cite{ghosh}. We
have assumed here (and also in the following) that the initial
wave packet is well below the dissociation limit, so that it
involves only the lower levels of HI molecule (the energy
distribution is peaked around the $\bar{n}=4$ vibrational level).
Figure~\ref{catsub}(a) shows the vibrational cat state after one
fourth of the fractional revival time. Here, the revival time is
$T_{rev}=4.89\times10^{4}$ a.u. Due to the anharmonicity of the
system, one can notice the different squeezing effects in the two
separated CSs forming the cat state. The number of ripples in the
interference region increases for increasing mean energy of the
initial CS. The sub-Planck scale structures appear in the
interference region at one eighth of the fractional revival time
[Fig.~\ref{catsub}(b)], where one has a coherent superposition of
four well distinct states, forming a so-called compass state
\cite{zurek2}. For this reason we shall focus our attention on the
effect of decoherence at this fractional revival time.

\begin{figure}[htpb]
\centering
\includegraphics[width=3.2in]{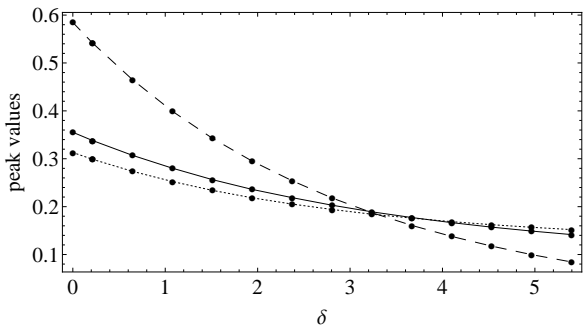}
\caption{Comparative variations between the left (dotted line) and
right (solid line) peaks and the central negative sub-Planck
region (dashed line) at $\frac{1}{8}T_{rev}$ with the coupling
parameter $\delta$ (in unit of $10^3$ a.u.). The environment
temperature is $T=10\hbar\omega_{01}/k_B$.}\label{peaks}
\end{figure}

In the case of a molecular vibration, a bosonic environment well
describes either the coupling via the dipole interaction with the
outside electromagnetic field or, in the case of a molecule
immersed in a liquid or gas, the coupling with the acoustic modes
of the solvent. In both cases the operator $\hat{O}$ is connected
with the position operator of our Morse oscillator. In fact,
$\hat{O}$ describes the upper triangular part (in the energy basis
representation) of the dipole moment operator of the molecule in
the electromagnetic case and of the vibrational coordinate $x$ in
the acoustic phonon bath case. However the two situations are
analogous because the dipole moment operator is proportional to
$x$. Both environments are super-Ohmic, that is, we have
\begin{equation}
2\pi\sigma^{2}(\omega)g(\omega)=\delta \omega^{3},
\end{equation}
with $\delta$ characterizing the strength of the
system-environment coupling. The physical meaning of the parameter
$\delta$ can be seen from the fact that the master equation
(\ref{meq}) implies that the relaxation rate from level $i$ to
level $j$, $\Gamma_{ij}$ at zero temperature is given by
\begin{equation}
\Gamma_{ij}=\delta {r_{ij}^{2}\omega_{ij}^{3}},
\end{equation}
where $r_{ij}$ is the corresponding matrix element of the position
operator between the two vibrational levels. Here we have chosen
the coupling constant $\delta$ such that the ratio
$\Gamma_{01}/\omega_{01}$ ranges from $0.5\times 10^{-5}$ to
$12.5\times10^{-5}$. These values correspond to reasonable values
of the coupling constant $\delta$; in fact, considering a typical
electric dipole moment of a diatomic molecule one gets $\delta
\simeq 10^{-12}$ m$^{-2}$ s$^{2}$ $=4.78\;a.u.$ for the case of an
electromagnetic environment. Instead, considering a dilute solvent
one gets $\delta \simeq 2\times 10^{-11}$ m$^{-2}$ s$^{2}$
$=95.66$ a.u. for the case of a phononic environment. One should
note that position variable in our study is a dimensionless
quantity (scaled by $r_0$). Hence, in the case of an
electromagnetic environment, the order of $\delta$ in our case
would be $\delta=4.37\times 10^3$ a.u., which is consistent with
our study (see Fig.~\ref{peaks}).
Moreover, the temperature of the environment is kept fixed at
$T=10\hbar\omega_{01}/k_{B}$. Figure~\ref{3ddeco} shows the Wigner
distribution at one eighth of the fractional revival time for
different values of the coupling with the bosonic environment.
Figure~\ref{3ddeco}(a) refers to no decoherence $(\delta=0$) and
therefore corresponds to Fig.~\ref{catsub}(b).
Figure~\ref{3ddeco}(b) instead corresponds to
$\delta=0.54\times10^{3}$ a.u. and Fig.~\ref{3ddeco}(c)
corresponds to a stronger decoherence, $\delta=2.2\times10^{3}$
a.u. One can clearly see that by increasing the coupling with the
bosonic environment, the interference region is more and more
affected.
\begin{figure}[htpb]
\centering
\includegraphics[width=3.2in]{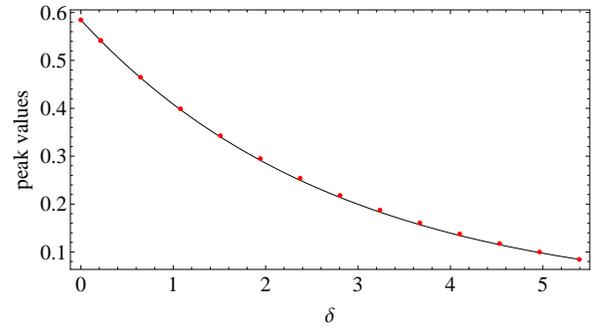}
\caption{Variation in the central negative sub-Planck region at
$\frac{1}{8}T_{rev}$ with the coupling parameter $\delta$ (in unit
of $10^3$ a.u.). Dots are the numerical data from our analysis. It
satisfies an exponential law (solid line) $A e^{-c \delta}$, with
$A=0.5847$ and $c=0.3585$. The environment temperature is $T=10
\hbar\omega_{01}/k_B$.}\label{exponential}
\end{figure}

\begin{figure*}[htpb]
\centering
\includegraphics[width=3.6in,height=2.in]{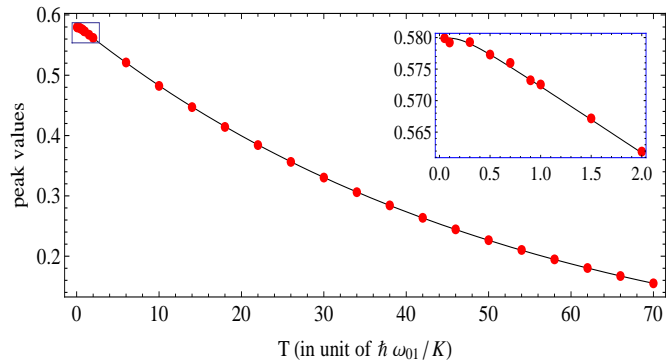}
\caption{(Color online) Variation in the central negative
sub-Planck region at time $\frac{1}{8}T_{rev}$ with the
environment temperature $T$ in the case $\delta=0.54\times10^{3}$
a.u. It follows a Bose-distribution law, $a
\exp\{-b/[e^{T_c/T}-1]\}$, for $a=0.5799$ and $b=0.0127$. Inset of
the figure shows the variation near the critical temperature
($T_c=0.6688$).}\label{temperature}
\end{figure*}

\begin{figure*}[htbp]
\centering
  \includegraphics[width=2.2in,height=1.8in]{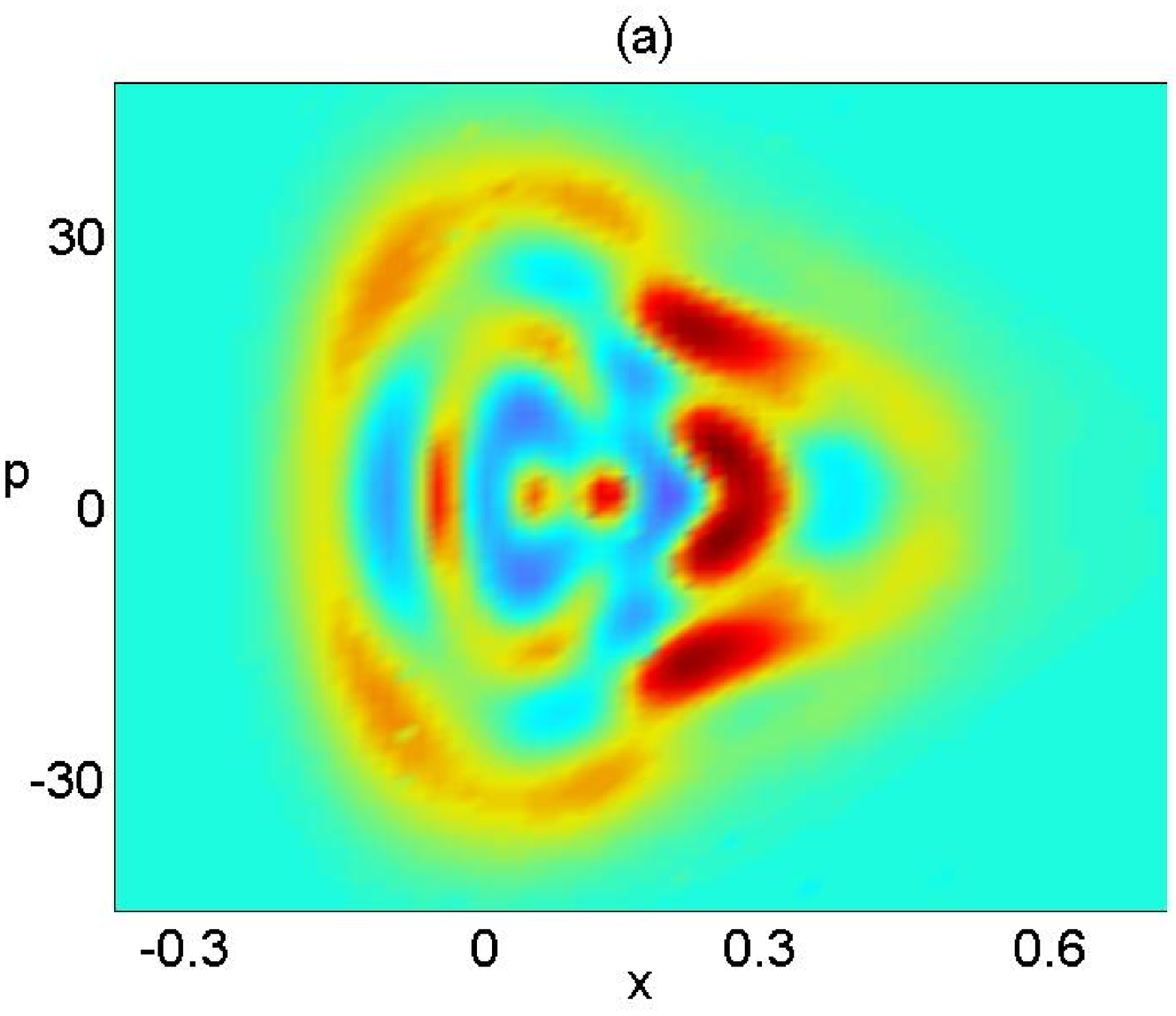}\!\!\!\!\!
   \includegraphics[width=2.4in,height=1.8in]{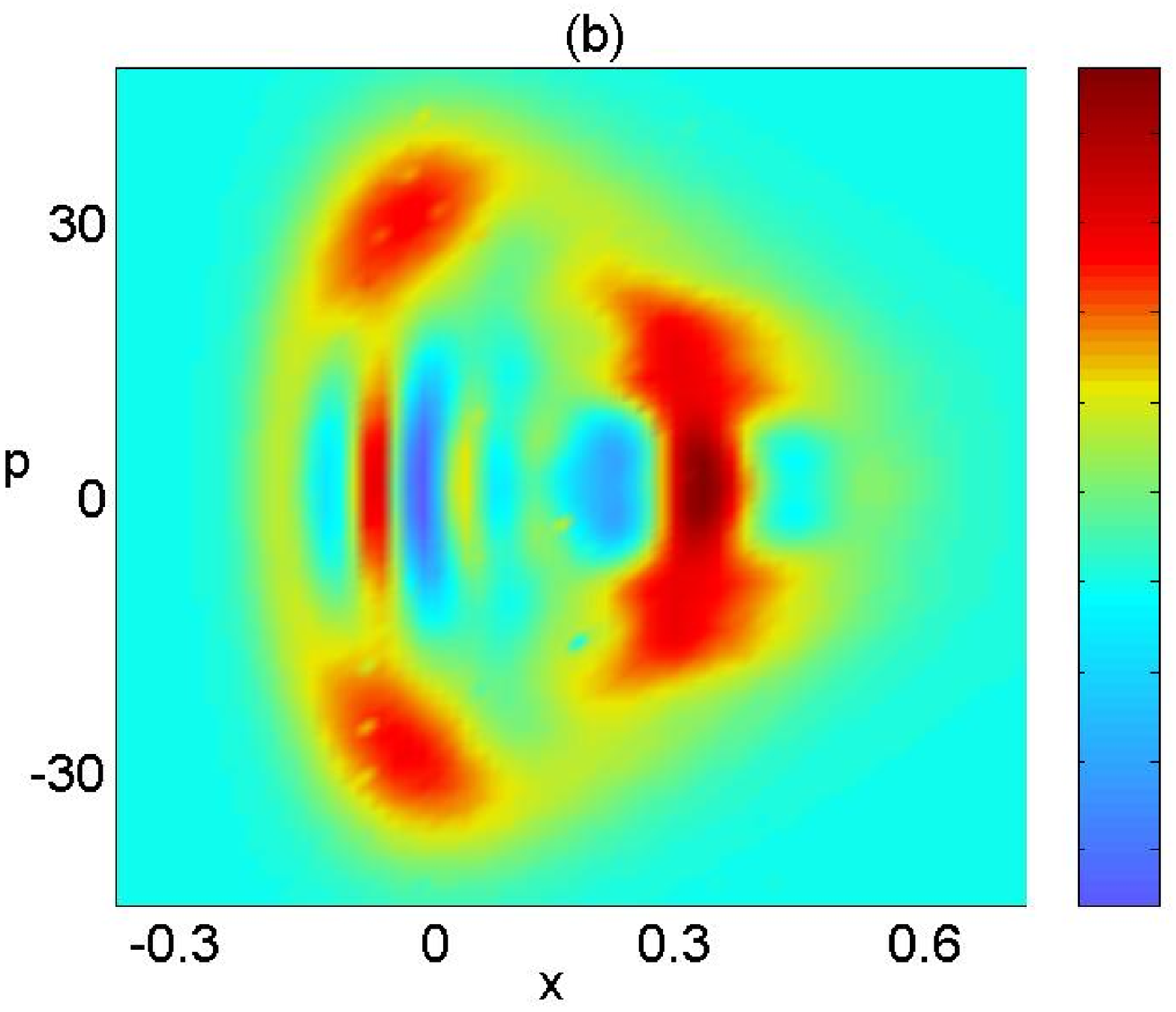}
\caption{(Color online) Wigner distribution at times (a)
$\frac{3}{8}T_{rev}$ and (b) $\frac{5}{8}T_{rev}$, with coupling
parameter $\delta=0.72\times10^{3}$ a.u. and temperature $T=10
\hbar\omega_{01}/k_B$. Central interference region in (a)
(sub-Planck scale structure) disappears at a later time in (b).
Here, $x$ and $p$ are the dimensionless position and momentum
variables, where $x=r/r_{0}-1$ and $p$ is the corresponding scaled
variable.}\label{3eight5}
\end{figure*}

As for the harmonic oscillator case \cite{zurek2,toscano},
decoherence affects the structure as a whole; also here for the
Morse oscillator, the sub-Planck structures due to quantum
interference are more affected than the individual isolated
coherent state components. In fact, a distinct difference can be
observed in the decay rate of the amplitude of the sub-Planck
scale structures and of the individual CSs. This is quantitatively
shown in Fig.~\ref{peaks}, where these decay rates are plotted
versus the decoherence strength $\delta$. We consider the left and
right peaks of the CSs at $p=0$ and a negative peak at $x=0.077$
and $p=-6.064$, appearing in the sub-Planck interference region.
The plot shows that the sub-Planck scale structure, i.e., the
central interference patterns (dashed line in Fig.~\ref{peaks}),
disappears earlier compared to the individual CSs, as it happens
in the harmonic case. It is possible to see that the decay of the
amplitude of the sub-Planck structure follows very well an
exponential law as a function of the decoherence strength
$\delta$, as expected in usual bosonic environments \cite{zurek1}.
We find that a linear exponential function $A e^{-c \delta}$ well
fits with our results, with $A=0.5847$ and $c=0.3585$.
Figure~\ref{exponential} shows how the rate of amplitude damping
of the chosen negative interference region (i.e., sub-Planck
region) matches with the exponential form.

It is now worth seeing the effect of environment temperature on
decoherence for a fixed value of the coupling constant $\delta$.
Owing to Eqs.~(\ref{elemat1}) and (\ref{elemat2}), one expects a
Bose-Einstein dependence on temperature of the decay of the
interference structures associated with sub-Planck structures, $a
\exp\{-b/[e^{T_c/T}-1]\}$, where $T_c$ corresponds to an effective
transition temperature below which the discrete structure of the
energy levels of the Morse oscillator starts to manifests itself.
This is confirmed by Fig.~\ref{temperature}, where the numerical
results for the value of the negative peak are plotted versus
temperature. The data, corresponding to $\delta=0.54\times10^{3}$
a.u., are well fitted by the above curve and the optimal fitting
parameters are $a=0.5799$, $b=0.0127$ and $T_c=0.6688$. The data
follow an exponential decay for $T/T_c\gg 1$, while the deviation
from the exponential law (associated with the Bose-Einstein
distribution dependence) is clearly visible only at very low
temperatures, $T< T_c$, in the magnified view in the inset of
Fig.~\ref{temperature}.

So far we have been studying the decoherence effect on the
sub-Planck scale structures at $1/8$ fractional revival time.
Hence, it is a natural question to ask what happens at larger
times when one can also obtain sub-Planck scale structures in the
interference region of four-way break up of a coherent state.
Thus, we extend our study to the four way break-up or the
decoherence through sub-Planck scale regions at $3/8$ and $5/8$
fractional revival times. One expects a larger influence of
decoherence on the sub-Planck structures for increasing times and
this is confirmed by Fig.~\ref{3eight5}. Interference fringes in
phase space are still visible at $3/8$ fractional revival time,
while in Fig.~\ref{3eight5}(b), corresponding to $5/8$ fractional
revival time, one can see that the sub-Planck structures
completely disappear due to the larger decoherence effect, whereas
the individual coherent states remain almost intact. The
environment temperature is kept constant at $T=10\hbar
\omega_{01}/k_B$, the coupling constant being
$\delta=0.72\times10^{3}$ a.u.

\section{Conclusions}
We have investigated the time evolution of a coherent state wave
packet in the Morse potential under the influence of a bosonic
environment describing either photonic or phononic excitations. We
have studied the effect of decoherence on the sub-Planck
structures in phase space by looking at the evolution of the
Wigner distribution. As it happens for the harmonic case,
sub-Planck scale structures come out as the most sensitive to
decoherence. A quantitative analysis provides an exponential decay
of the amplitude of the quantum interference structures as a
function of the coupling with the environment, in agreement with
usual predictions \cite{zurek1}. Influence of the environment
temperature on the decoherence is also shown quantitatively. This
is according to the Bose-distribution law. Longer time effect on
the decoherence is shown for providing another way to see the
sensitiveness of sub-Planck scale structures compare to their
original counterparts.

\end{document}